\documentclass[aps,prl,amsmath,showpacs,preprintnumbers,twocolumn,amssymb,amsmath,superscriptaddress]{revtex4}

\usepackage{graphicx}
\usepackage{mathptmx, textcomp}
\usepackage[latin1]{inputenc}
\usepackage{tabularx}
\usepackage{units}
\usepackage{color}

\begin{document}
\title{Reaching Fermi degeneracy via universal dipolar scattering}
%between identical particles
\affiliation{Institut f\"ur Experimentalphysik and Zentrum f\"ur Quantenphysik, Universit\"at Innsbruck, Technikerstra{\ss}e 25, 6020 Innsbruck, Austria}
\author{K. Aikawa}
\affiliation{Institut f\"ur Experimentalphysik and Zentrum f\"ur Quantenphysik, Universit\"at Innsbruck, Technikerstra{\ss}e 25, 6020 Innsbruck, Austria}
\author{A. Frisch}
\affiliation{Institut f\"ur Experimentalphysik and Zentrum f\"ur Quantenphysik, Universit\"at Innsbruck, Technikerstra{\ss}e 25, 6020 Innsbruck, Austria}
\author{M. Mark}
\affiliation{Institut f\"ur Experimentalphysik and Zentrum f\"ur Quantenphysik, Universit\"at Innsbruck, Technikerstra{\ss}e 25, 6020 Innsbruck, Austria}
\author{S. Baier}
\affiliation{Institut f\"ur Experimentalphysik and Zentrum f\"ur Quantenphysik, Universit\"at Innsbruck, Technikerstra{\ss}e 25, 6020 Innsbruck, Austria}
\author{R. Grimm}
\affiliation{Institut f\"ur Experimentalphysik and Zentrum f\"ur Quantenphysik, Universit\"at Innsbruck, Technikerstra{\ss}e 25, 6020 Innsbruck, Austria}
\affiliation{Institut f\"ur Quantenoptik und Quanteninformation,
 \"Osterreichische Akademie der Wissenschaften, 6020 Innsbruck, Austria}
\author{F. Ferlaino}
\affiliation{Institut f\"ur Experimentalphysik and Zentrum f\"ur Quantenphysik, Universit\"at Innsbruck, Technikerstra{\ss}e 25, 6020 Innsbruck, Austria}

\date{\today}


\begin{abstract}
We report on the creation of a degenerate dipolar Fermi gas of erbium atoms. We force  evaporative cooling in a fully spin-polarized sample down to temperatures as low as $0.2$ times the Fermi temperature. The strong magnetic dipole-dipole interaction enables elastic collisions between identical fermions even in the zero-energy limit. The measured elastic scattering cross section agrees well with the predictions from dipolar scattering theory, which follow a universal scaling law depending only on the dipole moment and on the atomic mass. Our approach to quantum degeneracy proceeds with very high cooling efficiency and provides large atomic densities, and it may be extended to various dipolar systems.
\end{abstract}

%This is a remarkable high number and it is among the highest cooling efficiency observed with fermonic atoms
\maketitle

Identical fermions with short-range interaction do not collide  at very low temperatures~\cite{Giorgini2008tou}. According to the rules of quantum mechanics, the requirement of anti-symmetry of the fermionic wave function causes the scattering cross section to vanish in the ultracold regime.  This makes ultracold fermions special in many respects. For instance, they realize perfectly non-interacting quantum systems, which can serve for sensitive interferometers~\cite{roati2004atom} and ultra-precise atomic clocks~\cite{nicholson2012comparison}. From another point of view, the absence of collisions means that direct evaporative cooling cannot work. 
%This method has been crucial in reaching quantum degeneracy in bosonic gases since the first realization of a Bose-Einstein condensates~\cite{davis1995bose,anderson1995observation}. 

The inapplicability of direct evaporative cooling  to fermions challenged scientists  to develop alternative strategies. The common solution is to use mixtures of two distinguishable atomic components~\cite{demarco1999onset}. In this scheme, fermions are sympathetically  cooled through elastic $s$-wave collisions with fermions in other spin states~\cite{demarco1999onset,granade2002all-optical,fukuhara2007degenerate,desalvo2010degenerate,taie2010realization}, with atoms belonging to a different isotope~\cite{Truscott2001oof,schreck2001quasipure,Mcnamara2006dbf,tey2010double,lu2012quantum}, or with atoms of a different chemical element~\cite{roati2002fermi,hadzibabic2002two-species,Silber2005qdm,Spiegelhalder2009cso}. 
%In all these cases, the collisional thermalization proceeds via $s$-wave scattering. 

The scenario is completely different in the presence of the long-range dipole-dipole interaction (DDI). While the effect of the short-range van der Waals interaction still freezes out at low temperatures, as it does for non-dipolar fermions, the DDI prevents the elastic cross section between identical fermions from vanishing. The corresponding Wigner threshold law, governing the threshold behavior of two-body scattering, gives a finite and energy-independent elastic cross section  \cite{Landau1977book, sadeghpour2000collisions, baranov2008theoretical}. As a key consequence, identical dipolar fermions can collide even in the zero-temperature limit.

%Its long-range nature drastically changes the threshold behavior of the two-body scattering with respect to the one of alkali and alkali-earth fermions, which is governed by the short-range van der Waals potential. At low energy, the collisional phase shift of two dipoles is linearly proportional to the scattering momentum and it is equal for all the partial waves \cite{Landau1977book, sadeghpour2000collisions, baranov2008theoretical}. 

%It follows that the threshold behavior of two-body scattering drastically change and that the Wigner threshold law gives a non-vanishing and energy-independent elastic cross section for dipolar fermions. As key consequence, identical fermions can collide even at zero temperature.

%While the short-range van der Waals interaction still freezes out at low temperature, as for ordinary alkali fermions, the long-range dipole-dipole interaction allows elastic scattering between identical fermions. 

%In spin-polarized fermions, the short-range van der Waals interaction freezes out at low temperature because of the Wigner threshold law  and only the long-range dipole-dipole interaction is responsible for the elastic scattering.  

Ultracold dipolar scattering is currently attracting a renewed interest in connection with recent experiments on polar molecules~\cite{ni2010dipolar,de2011controlling} and strongly magnetic atoms \cite{lu2011strongly,lu2012quantum,aikawa2012bose-einstein}. Early theoretical work on H atoms and atoms in electric fields suggested that dipolar scattering could provide an elastic cross section that is large enough for direct evaporative cooling of identical fermions~\cite{koelman1987spin-polarized, koelman1988lifetime,marinescu1998controlling,geist1999evaporative}.
Recent theoretical work has elucidated the universal character of the dipolar scattering~\cite{ticknor2008collisional,bohn2009quasi-universal,Idziaszek2010urc} and found that the elastic dipolar cross section is determined only by the mass and the dipole moment of the particles~\cite{bohn2009quasi-universal}. Recent experiments on fermionic ground-state polar KRb molecules have tested this prediction and have obtained evidence for the anisotropic character of the DDI~\cite{ni2010dipolar}. 
Experiments on using dipolar scattering for evaporative cooling have been reported for fermionic Dy~\cite{lu2012quantum} and KRb molecules~\cite{ye2013private}, both reaching temperatures on the order of the Fermi temperature $T_F$.

In this Letter, we report on the creation of a quantum degenerate dipolar Fermi gas of $^{167}$Er atoms.
 % realized by effiient dipolar-driven evaporative cooling between spin-polarized fermions  achieved by dipolar-driven evaporative cooling in a spin-polarized sample. 
 We demonstrate a  powerful approach  in which 
the underlying cooling mechanism relies  solely on dipolar scattering between spin-polarized fermions.
% and not on $s$-wave collisions between fermions in different spin states.
We observe a remarkably high cooling efficiency, leading to very dense Fermi gases with typically $6.4 \times 10^4$ atoms at a temperature of $T/T_F=0.2$ and a peak density of $4\times 10^{14}$ cm$^{-3}$.  %Remarkably, our approach  allows for very high atomic densities since inelastic collisions, occurring at short range, are inhibited by the $p$-wave centrifugal barrier, while elastic scattering between identical fermions is not suppressed at low temperatures because of the long-range dipole-dipole interaction (DDI).  These factors act unitedly to give high elastic collision rates and to provide very favorable good-to-bad-collision ratio, which is the ultimate limiting factor for efficient evaporative cooling (REF). We typically prepare degenerate Fermi gases with $7 \times 10^4$ atoms at a temperature of $T/T_F=0.2$, and at a density of $4\times 10^{14}$ cm$^{-3}$. Here, $T_F$ is the Fermi temperature. 
Finally, we confirm the prediction of the universal dipolar scattering theory~\cite{ticknor2008collisional, bohn2009quasi-universal} by measuring the Er elastic cross-section in  spin-polarized fermions via cross-dimensional thermalization~\cite{monroe1993measurement}. Our work opens up a conceptually novel  pathway to  quantum degeneracy in dipolar systems that can be generalized not only to other strongly magnetic atoms but also to ground-state polar molecules, for which the implementation of sympathetic cooling might be difficult. 

 %makes one of the highest cooling efficiency available on the market
 %We realize a very efficient evaporative cooling, giving a relatively large number of atoms, $7 \times 10^4$ atoms, at $T/T_F=0.21(1)$, where $T$ is the temperature of atoms and $T_F$ is the Fermi temperature. The density of the sample reaches $4\times 10^{14}$ cm$^{-3}$, which gives a high collision rate required for efficient evaporative cooling. 
%that originates from the large magnetic moment of 7 Bohr magneton of Er.
  %We determine the Er-Er elastic cross-section in a spin-polarized sample to be $3.1(3) \times 10^{-12}$~cm$^2$ by means of the cross-dimensional thermalization method~\cite{monroe1993measurement}. Our measurement  confirm the prediction of the universal dipolar scattering theory. 

%Our work has an important implication also for dipolar bosons, when DDI is larger than $s$-wave interaction between them. Efficient evaporative cooling is expected purely because of dipolar scattering.

%Experimental procedure:\\
%MOT:\\

The strong dipolar character of Er originates from its large magnetic moment $\mu$ of $7 \mu_B$, where $\mu_B$ is the Bohr magneton, and its large mass~\cite{baranov2008theoretical,baranov2012condensed}.
Among the six stable isotopes, Er has one fermionic isotope, $^{167}$Er,  with a large natural abundance of $23 \%$. While the  bosonic isotopes have no hyperfine structure, $^{167}$Er has a nuclear spin $I=7/2$, giving rise to a manifold of eight hyperfine levels and 104 magnetic sublevels in the electronic ground state~\cite{frisch2013hyperfine}. 
In spite of the much more complex energy structure of the fermionic isotope, our approach to quantum degeneracy is very similar to the one we have successfully used to condense the bosonic isotope $^{168}$Er~\cite{frisch2012narrow,aikawa2012bose-einstein}. It consists of a laser cooling stage followed by  direct evaporative cooling in an optical dipole trap (ODT). The fundamental difference with respect to the bosonic case is that the thermalization between spin-polarized fermions proceeds solely through dipolar elastic collisions. In the present work, we focus on spin-polarized fermions in the lowest hyperfine sublevel  $|F=19/2, m_F=-19/2\rangle$, where $F$ is the total spin quantum number and $m_F$ is its projection along the quantization axis.

\begin{figure}[t]
\includegraphics[width=1\columnwidth] {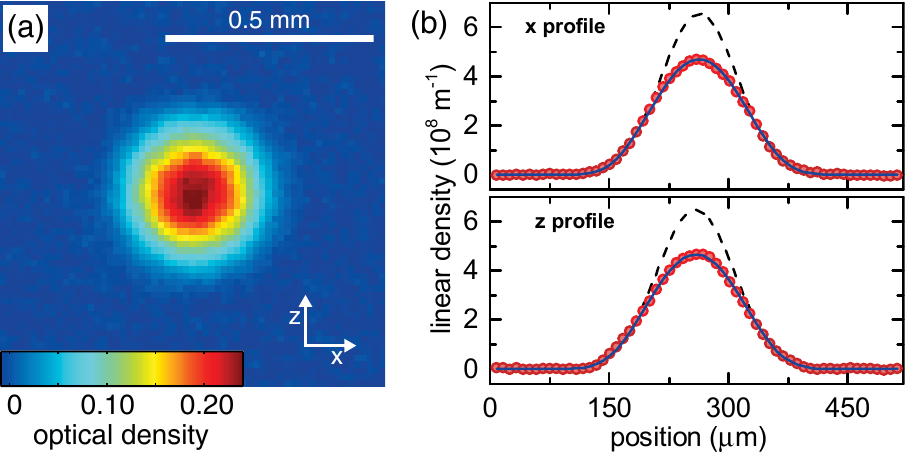}
\caption{(color online) Time-of-fight absorption image of a degenerate Fermi gas of Er atoms at $T/T_F = 0.21(1)$ after $t_{\rm TOF}=\unit[12]{ms}$ of expansion (a) and its density distribution integrated  along the $z$ direction (upper panel) and $x$ direction (lower panel) (b). 
The observed profiles (circles) are well described by fitting a  poly-logarithmic function to the data (solid lines), while they substantially deviate from a  fit using a Gaussian distribution to the outer wings of the cloud, i.\,e.\,$w$ (dashed lines). The absorption image is averaged over six individual measurements.}
\label{fig:fig_image}
\end{figure}

Our laser cooling scheme relies on a Zeeman slower operating at $\unit[401]{nm}$ and on a  magneto-optical trap (MOT) based on a narrow line at $\unit[583]{nm}$ \cite{frisch2012narrow}.  Both light fields act on transitions with quantum numbers $ F=19/2 \rightarrow F'=21/2~$, which are sufficiently closed for laser cooling. In our scheme, fermions in the MOT are naturally spin-polarized into the lowest magnetic sublevel $|19/2, -19/2\rangle$ because of a combined effect of gravity and the MOT light~\cite{frisch2012narrow}. We typically capture $1 \times 10^7$ atoms at $T = \unit[7]{\mu K}$ in the MOT. All measurements in the present work are performed by absorption imaging on the $401$-${\rm nm}$ transition. 

%of about $50 \%$.
%Dipole Trap:\\
%For evaporative cooling, we first transfer the atoms from the MOT  into a single-beam reservoir-like ODT at $\unit[1064]{nm}$ and then into a tightly-focused ODT at $\unit[1570]{nm}$. The reservoir trap is used as an intermediate step to increase the transfer efficiency from the MOT. It consists of a single horizontal beam with a power of $\unit[20]{W}$ and elliptic focus. The beam waists are approximately $\unit[20]{\mu m}$ and $\unit[200]{\mu m}$ in the vertical and horizontal direction, respectively~\footnote{ The ellipticity of the trap potential is chosen in order to maximize the transfer efficiency from the MOT and to the final ODT and it is tuned  in the horizontal plane by means of time-averaged optical potential\cite{friedman2001observation,milner2001optical}}.  From the reservoir trap, the atoms are loaded into a horizontal single-beam ODT at $\unit[1570]{nm}$ focused to a beam waist of  $\unit[15]{\mu m}$. The initial power of the $\unit[1570]{nm}$  beam is $ \unit[1.8]{ W}$, giving trap frequencies of  $(\nu_x, \nu_y, \nu_z) = \unit[(2147, 51, 2316)]{Hz}$, where $z$ is the direction of gravity. At this stage, we have $1.5 \times 10^6$ atoms at $\unit[28]{\mu K}$, corresponding to $T/T_F = 4.4$. 
For evaporative cooling, we first transfer the atoms from the MOT  into a single-beam large-volume ODT at $\unit[1064]{nm}$ and then into a tightly focused ODT at $\unit[1570]{nm}$. The first trap is used as an intermediate step to increase the transfer efficiency from the MOT. It consists of a single horizontal beam with a power of $\unit[20]{W}$ and elliptical focus. The beam waists are approximately $\unit[20]{\mu m}$ and $\unit[200]{\mu m}$ in the vertical and horizontal direction, respectively.  The corresponding trap depth is roughly  $ \unit[100]{\mu K}$. From the large-volume trap, the atoms are loaded into a tightly focused ODT at $\unit[1570]{nm}$. This second trap is made of a single horizontal beam, which is collinear  to the large-volume trapping beam  and has a waist of  $\unit[15]{\mu m}$.  The initial power of the $\unit[1570]{nm}$  beam is $ \unit[1.8]{W}$, corresponding to trap frequencies of  $(\nu_x, \nu_y, \nu_z) = \unit[(2147, 51, 2316)]{Hz}$ and a trap depth of about $k_B\times \unit[190]{\mu K}$. Here, $z$ is the direction of gravity. At this stage, we have $1.5 \times 10^6$ atoms at $T/T_F = 4.4$ with $T=\unit[28]{\mu K}$ and a peak density of about $1.2\times 10^{14}$ cm$^{-3}$. The Fermi temperature is defined as $T_F =  h \bar\nu (6N)^{1/3}/k_B$, where  $\bar\nu$ is the geometric mean of the trap frequencies and $h$ is the Planck  constant.
%The corresponding trap depth is XX and XX.
%
% Evaporative cooling
We force evaporation by reducing the power of the horizontal beam in a near-exponential manner. 
%The full sequence for evaporative cooling  is optimized by minimizing $T/T_F$ at each stage. 
When $T_F$ is reached, we introduce a vertical beam at $\unit[1570]{nm}$ to confine the fermions into the crossed region created by the two beams and to preserve the atomic density. Its power is gradually increased and reaches $\unit[1.2]{W}$ at the end of the evaporation. The vertical beam has a beam waist of $\unit[33]{\mu m}$.   %1.2W 
%procedure, we turn on a $\unit[1570]{nm}$ vertical beam intersecting with the horizontal beam to confine atoms into the crossed region. The beam waist of the vertical beam is 33$\mu$m. Our final trap has frequencies of $\unit[(470, 346, 345)]{Hz}$. 
During evaporation, we apply a homogeneous guiding magnetic field to maintain the spin-polarization in the system. At high temperature, the magnetic field value is about $\unit[1.7]{G}$, which is large enough to avoid any thermal excitation into higher spin states. For temperature below $3.2 T_F$, we  decrease the value of the magnetic field to $\unit[0.59]{G}$, where we observe a slightly better evaporation efficiency. 
%We observe 16 $p$-wave Feshbach resonances below $\unit[1]{G}$, with typical widths of around 10 mG at $\unit[1]{\mu K}$. The magnetic fields are chosen to avoid these Feshbach resonances. 
After $\unit[10]{s}$ of forced evaporation, we obtain a deeply degenerate Fermi gas.

Figure \ref{fig:fig_image}(a) shows a typical time-of-flight (TOF) absorption image of a  degenerate dipolar Fermi gas of $N=6.4 \times 10^4$ and a peak density of $n_0=4\times 10^{14}$ cm$^{-3}$ at  $T/T_F = 0.21(1)$ with $T_F= \unit[1.33(2)]{ \mu K}$. At this point, our trap frequencies are $\unit[(470, 346, 345)]{Hz}$. 
Fermi degeneracy reveals itself in a smooth change of the momentum distribution from a Maxwell-Boltzmann to a Fermi-Dirac distribution \cite{Inguscio2006ufg}. Correspondingly, the atomic density profile is expected to change its Gaussian shape into a poly-logarithmic one. A fit to TOF images reveals that  at temperatures above $\approx 0.5T_F$ the Gaussian and poly-logarithmic function are hardly distinguishable from each other and both describe the data well. By further decreasing the temperature, we observe a gradually increasing deviation from the Gaussian shape. This deviation is evident in Fig.\,\ref{fig:fig_image}(b), which shows a density profile at $T/T_F = 0.21(1)$. A Gaussian fit to the outer wings of the cloud, i.\,e.\,outside the disk with radius $w$, with $w$ being the $1/e$ diameter of the Gaussian fit to the entire cloud,  clearly overestimates the population at the center of the cloud. This is a fingerprint of Fermi degeneracy, meaning that the population of low-energy levels is limited by the Pauli  exclusion principle.

%the atomic density profile. At high $T$, the profile has a Gaussian shape, coming from the  Maxwell-Boltzmann distribution, while a $T<T_F$ it starts to acquire a poly-logarithmic form, reflecting the  Fermi-Dirac distribution, which is typical of a degenerate Fermi gas  \cite{Inguscio2006ufg}. 

  %At this degree of degeneracy, the density profile is accurately described by the Fermi-Dirac distribution, while a Gaussian fit to the outer wings of the cloud clearly overestimates the population at the center of the cloud;
%see Fig.\,\ref{fig:fig_image}(b). This is a typical fingerprint of Fermi degeneracy, meaning that low-energy levels are well occupied and their population is limited by the Pauli  exclusion principle. 

%%As expected, the density profile shows a substantial deviation from the Gaussian fit, while it is accurately described by the Fermi-Dirac distribution; see Fig.\,\ref{fig:fig_image}(b). 
%a two-dimensional polylogarithmic fit resulting from 
%At the classical regime of $T \geq 0.5 T_F$, the distribution is hardly distinguishable from a Gaussian function and thus the TOF images are fitted by a Gaussian function reflecting the Maxwell-Boltzmann distribution. 
%By contrast, at the quantum degenerate regime of $T < 0.5 T_F$, the distribution shows a sizable deviation from a Gaussian function and thus the TOF image is fitted by a two-dimensional polylogarithmic function reflecting the Fermi-Dirac distribution~\cite{desalvo2010degenerate,tey2010double}. 
\begin{figure}[t]
\includegraphics[width=1\columnwidth] {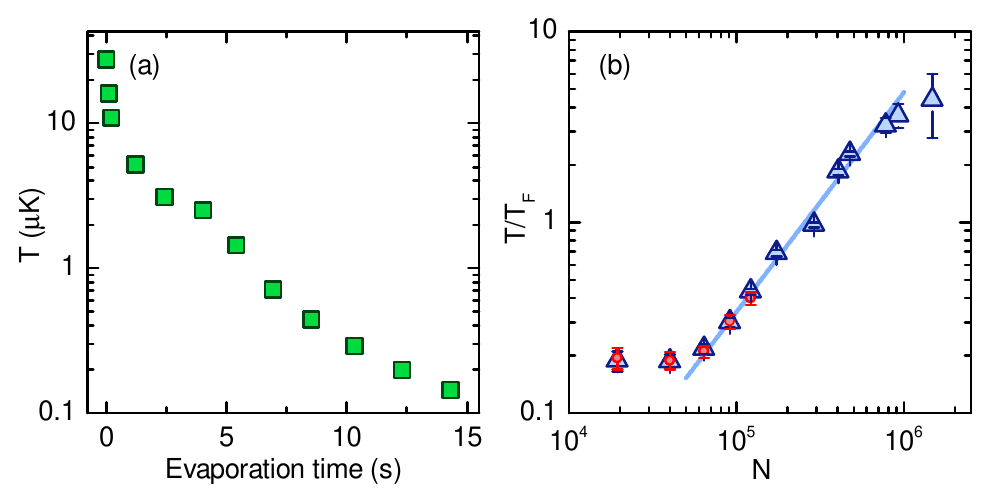}
\caption{ (color online) Evaporation trajectory to Fermi degeneracy. (a) Temperature evolution during the evaporation ramp and (b) corresponding $T/T_F$ versus N. The ratio $T/T_F$ is obtained from the width $\sigma$ of the distribution (triangles) and from the fugacity (circles); see text.
%Fermi-Dirac fits to the atomic density profiles, where either the width (triangles) or the fugacity (circles) is used as a fit parameter. 
The error bars  originate from statistical uncertainties in temperature, number of atoms, and trap frequencies for the width measurements and the standard deviations obtained from several independent measurements for the fugacity. 
The solid line is a linear fit to the data for  $0.2 < T/T_F < 4$.}
\label{fig:fig_evap}
\end{figure}
%$T/T_F$ is plotted as a function of the atom number.
%The entire evaporation procedure has a duration of $\unit[14]{ s}$. 
%The ratio  $T/T_F$ controls the degree of degeneracy of the Fermi gas. 
In all our measurements, we extract $T/T_F$ from fits to the density profiles  using either a poly-logarithmic or a Gaussian function.  
In the former case,  the fit gives both the fugacity $\zeta$ and the parameter $\sigma$ characterizing the width of the distribution. The fugacity directly gives  $T/T_F=[-6\times {\rm Li}_3(- \zeta)]^{-1/3}$, with ${\rm Li}_n$ being the $n$-th order poly-logarithmic function \cite{desalvo2010degenerate,tey2010double}. 
The parameter  $\sigma$ is related to the atomic temperature by $T=m \sigma^2/(k_B t^2_{\rm TOF})$, where $t_{\rm TOF}$ is the time of flight and $m$ is the mass of  $^{167}$Er, and together with $T_F$,  calculated from $N$ and $\bar\nu$, gives a more indirect value for $T/T_F$. We determine $T/T_F$ using both methods, which show well consistent results. 

To get deeper insights into the evaporation process and the underlying collisional properties we study the evaporation trajectory.
%We study the evaporation trajectory  by measuring the evolution of $T/T_F$  as a function of $N$. 
Figure \ref{fig:fig_evap} summarizes our results. We observe that the evaporation first proceeds with high  efficiency down to temperatures well below $T_F$  and then plateaus at about  $T/T_F=0.2$. The latter behavior suggests that further cooling is limited by Pauli blocking   \cite{demarco1999onset,desalvo2010degenerate,tey2010double,fukuhara2007degenerate} and that more thoroughly optimized evaporation ramps might be needed to reach even lower temperatures.  The deepest degeneracy we attained is $T/T_F = 0.19(1)$ with $N=4.0 \times 10^4$.  
%%% AAAAA%%%%%%The key feature of our scheme is that the underlying cooling mechanism to quantum degeneracy relies  only on dipolar scattering between spin-polarized fermions and not on $s$-wave collisions between fermions in different spin states.
From the slope of the evaporation trajectory, we obtain the efficiency parameter $\gamma$.
This parameter quantifies the gain in phase-space density ${\rm PSD}$ at the expense of the atom number and can be written as $\gamma = -d({\rm \ln PSD})/d({\ln} N) = -3 \times d({\ln}T/T_F)/d({\ln} N)$.
From a linear fit to the data down to $T/T_F=0.2$, we find $\gamma = 3.5(2)$.  
This remarkably large number is in the league of the best evaporation efficiencies observed in experiments with ultracold atoms based on  $s$-wave scattering, including our experiments with the bosonic $^{168}$Er~\cite{aikawa2012bose-einstein} and experiments on strongly interacting two-component Fermi gases~\cite{granade2002all-optical,bartenstein2004crossover,zwierlein2004condensation}.

\begin{figure}[t]
\includegraphics[width=1\columnwidth] {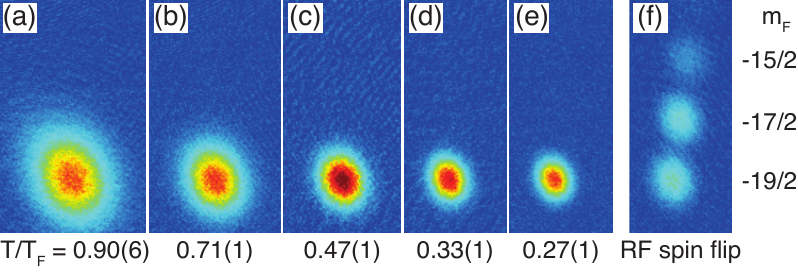}
\caption{(color online) Absorption images of the atomic cloud with a Stern-Gerlach separation of the spin components. A magnetic field gradient of about 40 G/cm is applied during the expansion for about $\unit[7]{ms}$. (a)-(e) Along the entire evaporative cooling sequence, atoms are always spin-polarized in the lowest hyperfine sublevel $|F=19/2, m_F=-19/2\rangle$. $T/T_F$ of the atomic samples are indicated in each panel. In (f) the image is obtained right after RF mixing of the spin states for the sample at $T/T_F=0.33(1)$. The three clouds correspond to the magnetic sublevels  $m_F=-19/2$, $ -17/2$, and $-15/2$ from bottom to top. }
\label{fig:fig_sterngerlach}
\end{figure}

Our interpretation of the cooling process in terms of dipolar scattering relies on the full spin polarization of the sample.  Another spin state being present would lead to s-wave collisions in the sample. Therefore it is important to make sure that we do not have any other spin state present. For this reason, we carry out a dedicated set of Stern-Gerlarch-type measurements at various stages of the evaporation. 
During the whole evaporation sequence, we never observe any population in spin states different from the $m_F=-19/2$ state. 
Figure \ref{fig:fig_sterngerlach}(a-e) shows the relevant portion of the TOF image, where atoms are observed. 
To identify unambiguously the spatial positions of the different spin components, we intentionally prepare a spin mixture by radio-frequency (RF) transfer; see Fig.\,\ref{fig:fig_sterngerlach}(f).
It is worth to mention that we observe fast spin relaxation when a multi-component mixture is prepared~\cite{aikawainprep}.

\begin{figure}[t]
\includegraphics[width=1\columnwidth] {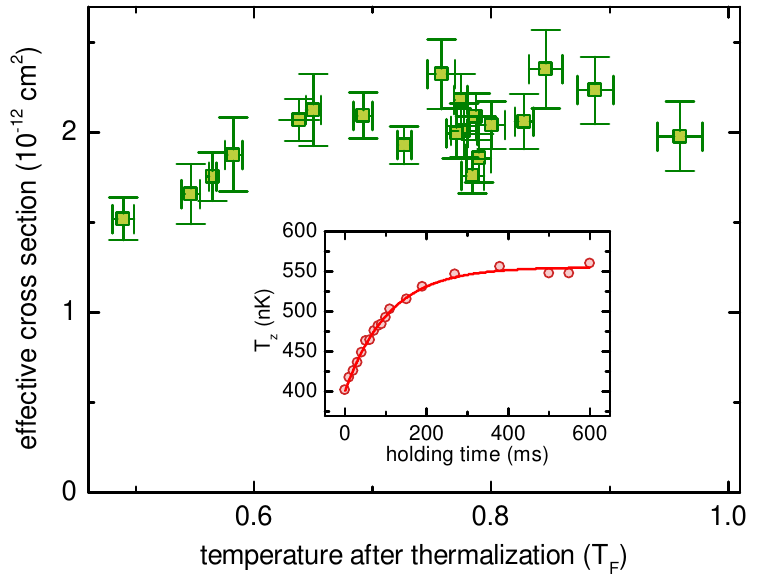}
\caption{(color online) Effective elastic cross-section as a function of $T/T_F$ after thermalization. %The effective elastic cross-section is derived by the cross-dimensional thermalization method, assuming that all the atoms are contributing to the thermalization process. In  the degenerate regime, a decrease in the effective cross-section indicates that the Pauli blocking of collisions begins to play an important role. 
In the non-degenerate regime, the effective cross section is constant and gives a mean value of $2.0(5) \times 10^{-12} {\rm cm}^2$. The error bars for each point contain the statistical uncertainties of the time constant for  cross-dimensional thermalization, of the trap frequencies, and of the temperature. A typical cross-dimensional thermalization measurement with an exponential fit to the data is shown in the inset. $T_z$ is the temperature along the $z$-direction. }
\label{fig:fig_crosssection}
\end{figure}

The  effectiveness of our evaporative cooling scheme suggests a very favorable ratio of the elastic scattering rate to the inelastic one. 
We explore elastic scattering by measuring the elastic dipolar cross section $\sigma_{\rm el}$ in our spin-polarized fermionic sample via cross-dimensional thermalization experiments~\cite{monroe1993measurement}. We compress the system in one spatial direction by increasing the power of the vertical beam by about a factor of three. We then monitor the time evolution of the temperature in the other direction, as shown in the inset of Fig.~\ref{fig:fig_crosssection}.  The time constant $\tau $ for cross-dimensional thermalization is directly connected to  $\sigma_{\rm el}$ through the relation $\tau = \alpha / (\bar{n} \sigma_{\rm el} v)$, where $\alpha$ is the number of collisions required to thermalize, $\bar{n}$ is the mean density, and $v = 4 \sqrt{k_B T/(\pi m)}$ is the mean relative velocity. A delicate point of our analysis is the estimation of $\alpha$, which depends on the underlying scattering mechanism. We employ $\alpha = 4.1$, which has been numerically calculated for non-dipolar $p$-wave collisions and has been applied to KRb polar molecules~\cite{ni2010dipolar}. Although $p$-wave collisions are expected to be the leading term in dipolar scattering of identical fermions, more detailed calculations of $\alpha$ might be needed to fully account for the mixing of partial waves resulting from the DDI~\cite{privateBohnJin}.

In this way, we explore elastic scattering over a wide range of atom numbers  from $3\times 10^4$ to $1.1 \times 10^5$ and for various final temperatures ranging from $300$ to $600$ nK. Our findings at $\unit[0.59]{G}$~\cite{note6} are shown  in Fig.\,\ref{fig:fig_crosssection}. In the non-degenerate regime ($T\gtrsim T_F$), we obtain a constant elastic cross section
with a mean value of  $\unit[2.0(5)\times 10^{-12}]{cm^2}$, corresponding to $[2.7(3)\times 10^2 a_0]^2$, where $a_0$ is the Bohr radius. The error bar is mainly due to systematic uncertainties in trap frequencies, temperature, and number of atoms. 
%Here, the error bar contains the standard deviation and the statistical uncertainties from the trap frequencies and the temperature measurements. This value is comparable to (give example of Na for instance).
Below $T_F$, the effect of quantum degeneracy becomes visible through a suppression of scattering events caused by Pauli blocking. In this regime, we can interpret our measurements in terms of an effective elastic cross section, which also includes the Pauli suppression factor. As expected, we observe a substantial decrease of the effective $\sigma_{\rm el}$  for decreasing $T/T_F$, similarly  to  the case of  $s$-wave collisions between fermions in different spin states~\cite{Demarco2001pbo}.

%The key feature of our scheme is that the underlying cooling mechanism to quantum degeneracy relies  only on dipolar scattering between spin-polarized fermions and not on $s$-wave collisions between fermions in different spin states.
%We confirm the spin purity of our system by performing Stern-Gerlarch-type experiments at each stage of the evaporation; see Fig.\,\ref{fig:fig_sterngerlach}(a-e). As shown in the figures, we always observe the fermions to be fully polarized in the  $m_F=-19/2$ level. To pinpoint the spatial positions of the spin components, we intentionally prepare a spin mixture by applying  a radio-frequency signal to transfer the atoms into higher spin states; see Fig.\,\ref{fig:fig_sterngerlach}(f). 

%Neglecting the effect of Pauli blocking,
Dipolar scattering theories predict  an energy-independent elastic cross section for identical fermions in the low-energy regime \cite{Landau1977book, sadeghpour2000collisions, baranov2008theoretical}. The cross section is predicted to follow a universal scaling law that is fully determined by a single parameter   - the dipolar length ${D}$~\cite{bohn2009quasi-universal}  - and it reads as
\begin{equation}
\sigma_{\rm el}= 6.702 \times {D}^2,
\end{equation}
where  ${D}= 2 \pi^2 d^2 m/h^2$ with $d^2=\mu_0 \mu^2/ (4\pi)$ and $\mu_0$ being the vacuum permeability. This equation shows a clear analogy to the ordinary $s$-wave scattering, where $D$ plays the role of the scattering length. For the Er parameters, the universal theory predicts  $\sigma_{\rm el}=\unit[1.8\times 10^{-12}]{ cm^2}$, which is in  reasonable agreement with the measured value.  The small deviation might be due to the chosen value for $\alpha$, to systematic errors, or to a residual effect of the short-range physics, which is not included in the theory.

Our observations suggest that inelastic losses are very weak. Since the atoms are fully polarized in the lowest spin state, inelastic losses can only be caused by collisions with the background gas and by three-body decay. To investigate this more quantitatively, we carry out atom-decay measurements by recording the number of atoms as a function of the hold time in an ODT initially loaded with $N \simeq 1\times 10^5$ atoms at $T/T_F \simeq 0.47$. In spite of the very high peak density of $3 \times 10^{14} {\rm cm}^{-3}$, we find the atom number to decay in a purely exponential manner (time constant $\unit[40]{s}$) without showing any signature of three-body processes. From this observation we can derive an upper limit for the three-body recombination rate constant as low as $L_3 \leq 3 \times 10^{-30}$ cm$^6$/s.

The remarkable efficiency of evaporative cooling in a single-component Fermi gas of Er and the exceptionally high densities together with low inelastic collision rates can be understood in terms of a very favorable combination of the DDI with the $p$-wave barrier. While DDI is strong enough to provide us with a sufficient cross section for elastic collisions, it is weak enough to preserve a substantial repulsive barrier for any alignment of the colliding dipoles. Even for the case of maximum dipolar attraction (head to tail configuration), the effective potential, given by the interplay between the $p$-wave barrier and the DDI, features a repulsive barrier with a maximum height $V(r_{\rm max})=2\hbar^2/(27 m D^2)$ at $r_{\rm max}=3D$. For Er, the barrier height still exceeds $k_B \times 7\,\mu$K, which is much larger than all collision energies in the final evaporation stage. This prevents atoms from getting close to each other and three-body decay, which requires short-range interactions, is strongly suppressed.

In conclusion, we produce a degenerate dipolar Fermi gas of $^{167}$Er atoms. We demonstrate direct evaporative cooling of identical fermions via universal dipolar scattering. Our method provides two key advantages: feeble inelastic losses and exceptionally high attainable densities. The former aspect is favorable for reaching low values of $T/T_F$, which are ultimately limited by the so-called hole-heating mechanism caused by inelastic losses~\cite{timmermans2001degenerate,carr2004limits}. The latter aspect has important consequences for dipolar physics. The relevant energy scale for dipolar phenomena at the many-body level is given by $n_0 d^2$~\cite{baranov2008theoretical,baranov2012condensed}. Given the high densities achieved here, our degenerate Fermi gas of Er currently is the most dipolar quantum gas available in experiments, with $n_0 d^2$ being $0.92\%$ of the Fermi energy. We speculate that even much higher densities than the ones here attained may be achieved since we do not see any limiting process. This may open a way for observing $p$-wave pairing in dipolar gases and for the creation of an anisotropic Fermi superfluid~\cite{you1999prospects,baranov2002superfluid}.

\begin{acknowledgments}
We are grateful to J.\,Bohn, C.\,Salomon, M.\,Baranov, and M.\,Zwierlein for fruitful discussions. We thank M. Springer for technical support and NKT Photonics for lending us the 1570 nm fiber laser. This work is supported by the Austrian Ministry of Science and Research (BMWF) and the Austrian Science Fund (FWF) through a START grant under Project No. Y479-N20 and by the European Research Council under Project No. 259435. K. A. is supported within the Lise-Meitner program of the FWF.
\end{acknowledgments}

\bibliographystyle{apsrev}

%\bibliography{Er_DFG_ref,ultracold}

\end{document}